\documentclass[aps,prb,twocolumn,groupedaddress,amsfonts,amssymb,amsmath]{revtex4}

\usepackage{bm}
\usepackage{graphicx}

\begin{document}
\title{Elastic Tensor of YNi$_2$B$_2$C}
\author{P.M.C. Rourke}
\author{Johnpierre Paglione}
\author{F. Ronning}
\author{Louis Taillefer}
\affiliation{Department of Physics, University of Toronto, Toronto, Ontario, M5S 1A7, Canada}
\author{K. Kadowaki}
\affiliation{Institute of Materials Science, University of Tsukuba, Tsukuba, Ibaraki 3058573, Japan}

\date{\today}

\begin{abstract}
The complete elastic tensor of YNi$_2$B$_2$C was determined by application of the resonant ultrasound spectroscopy technique to a single-crystal sample. Elastic constants were found to be in good agreement with partial results obtained from `time-of-flight' measurements performed on samples cut from the same ingot. From the measured constants, the bulk modulus and Debye temperature are calculated.
\end{abstract}
\maketitle

The discovery of superconductivity in the borocarbide quaternary intermetallic compounds\cite{Cava} has generated a great deal of interest due to the observed interplay between superconductivity and magnetic order. Compounds in this family have formulae of the form \emph{R}Ni$_2$B$_2$C, where choice of the rare earth ion, \emph{R}, strongly affects the physical properties, which range from superconducting with no magnetic ordering (\emph{R}=Y) to non-superconducting with a N\'{e}el temperature of 19~K (\emph{R}=Gd).\cite{Canfield}

Recent heat transport measurements on non-magnetic borocarbides suggest that the superconductivity is very anomalous, with a gap that is either highly anisotropic\cite{Boaknin} or has point nodes.\cite{Matsuda} To elucidate the nature of the order parameter, a powerful tool is ultrasound attenuation (see for example Ref.~\onlinecite{Lupien_PE}). As a first step towards measuring the attenuation of sound it is important to establish the full elastic tensor of the material.

The 2$^{nd}$ order elastic constants relate stresses placed on a material along various directions to the resulting deformations of the crystal lattice. As such, knowledge of a compound's elastic constants can be useful when investigating a number of solid state properties, such as interatomic potentials, equations of state, phonon spectra, the Debye temperature, and the bulk modulus. In the most general case, a highly anisotropic solid can have 21 independent second-order elastic constants, but the crystal structure for all compounds in the \emph{R}Ni$_2$B$_2$C group is body-centered-tetragonal (space group \emph{I}4/\emph{mmm}), consisting of \emph{R}-C planes separated by Ni$_2$B$_2$ layers stacked along the \emph{c} axis,\cite{Lynn} reducing the number of independent constants to 6.\cite{Migliori} In Voigt notation, these constants are denoted \emph{C}$_1$$_1$, \emph{C}$_3$$_3$, \emph{C}$_1$$_3$, \emph{C}$_1$$_2$, \emph{C}$_4$$_4$, and \emph{C}$_6$$_6$; all other tensor entries can be retrieved from these six.

In this article, the elastic constants of YNi$_2$B$_2$C are determined at room temperature via two methods: resonant ultrasound spectroscopy and ultrasonic time-of-flight. Resonant ultrasound spectroscopy (RUS)\cite{Migliori} allows the determination of the complete elastic tensor of a material in one measurement, and was used as the primary experimental technique of this study. As a complement to the RUS measurements, elastic constants were also determined by measuring sound velocities using the more conventional `time-of-flight' method. Note that this study is not simply a confirmation of existing measurements: it adds a missing elastic constant and provides a very accurate second measure to secure time-of-flight results in slight disagreement.

\begin{figure}
 \centering
 \includegraphics[width=3in]{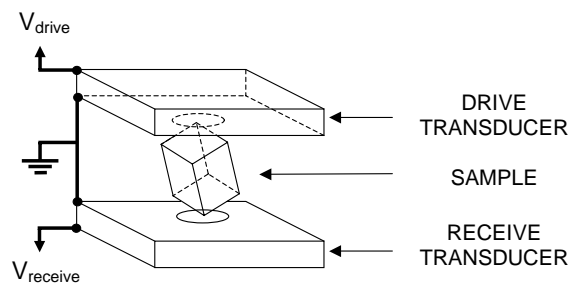}
 \caption{\label{fig:rus} Schematic of a typical RUS experimental set-up (from Ref.~\onlinecite{Paglione_RUS}).}
\end{figure}

The RUS method utilizes the normal modes of vibration of a solid sample of known density, geometry and crystal structure to determine the elastic tensor. To perform an RUS measurement, the normal mode frequency spectrum of the sample in question is first obtained. This is done by holding the sample between two piezo-electric transducers, generally by its corners in order to reduce damping of the vibrational modes. One transducer is driven at a frequency which is swept across the region of the spectrum containing the normal modes, thus driving the sample resonances, while the other picks up the resulting amplitudes of crystal vibration (see Fig.~\ref{fig:rus}).

Once a list of normal mode frequencies has been compiled, the elastic constants can be calculated. While it is possible to directly compute the normal mode frequencies of a sample, given the dimensions, mass and elastic constants, there is no known analytic method to do the reverse, so a computerized fitting algorithm has been developed\cite{Migliori} to numerically complete the calculation. This algorithm changes the elastic constants, calculates resulting normal mode frequencies, compares these calculated frequencies to the experimentally measured list and then iteratively repeats the process until the best fit to the data is obtained.

In the time-of-flight method, short ultrasound pulses are propagated through a sample carefully prepared with two flat, parallel faces, and later detected in order to obtain the sound velocity in the direction of pulse propagation. The diagonal entries of the elastic tensor can be easily calculated using the velocities measured along various principal crystal directions. The off-diagonal constants, however, require measurements of sound velocities in non-principal directions, which can be difficult. The equation (\emph{C}$_{11}$ + \emph{C}$_{12}$ + 2\emph{C}$_{66}$)/2 = $\rho$ $v^{2}$$_{[110/L]}$, where $\rho$ is the mass density of the material, and $v_{[110/L]}$ is the velocity of sound propagating in the [110] direction with a longitudinal polarization, is an example of how a sound velocity is related to the elastic constants for a tetragonal crystal.\cite{Truell} In this notation scheme, a transverse polarization, for example in the [010] direction, is indicated by $T010$.

With the complete elastic tensor, the bulk modulus of a tetragonal crystal, $B_0=-V\left(\frac{dP}{dV}\right)$, is easily obtained via the equation:
\begin{equation} \label{1:1}
B_0 = \frac{C_{33}(C_{11} + C_{12}) - 2C_{13}^2}{C_{11} + C_{12} + 2C_{33} - 4C_{13}}
\end{equation}

Also, an estimate of the Debye temperature can be made. For a crystal with tetragonal structure the following equation\cite{Alers} is employed:
\begin{equation} \label{1:2}
\theta _{D} = \frac{h}{k_{B}} \left(\frac{9N}{4\pi V}\right)^{-1/3} \rho^{-1/2} J^{-1/2},
\end{equation}
wherein \emph{h} is Planck's constant, \emph{k$_B$} is Boltzmann's constant, \emph{N} is the number of atoms per unit cell, \emph{V} is the volume of the unit cell, $\rho$ is the mass density of the material, and \emph{J} is the harmonic series expansion of an integration over sound velocities in all directions, written in terms of the elastic constants (Eqs. 32 and 33 in Ref.~\onlinecite{Alers}).

A large high-quality single crystal specimen of YNi$_2$$^{11}$B$_2$C was grown using the floating zone method. From this specimen a sample was cut using spark erosion and then polished into the shape of a rectangular parallelepiped, with sides aligned along the principal crystal axes as determined by Laue x-ray backscattering measurements.

Dimensions of the polished sample were measured to be 0.0633(8) cm $\times$ 0.0873(9) cm $\times$ 0.238(2) cm. The sample mass was 0.00832(3) g, yielding a density of 6.3(1) g/cm$^3$. Note that this is somewhat different from the density of 6.09(1) g/cm$^3$ for YNi$_2$$^{11}$B$_2$C calculated using lattice constants obtained from x-ray/neutron measurements.\cite{Lynn,Chakoumakos,Yelon,Buchgeister,Jaenicke-Roessler,Ignatov}

A number of other measurements do indicate good sample quality. Specifically, a residual resistivity ratio (RRR) of 14.2, a superconducting width of 0.3~K, and a $T_c$ of 14.7~K, were observed via resistivity measurements on our specimen, which compare favourably with published values.\cite{Cava,Chakoumakos,Yelon,Ignatov,Rathnayaka,Hong,Godart,Isida,Andreone,Movshovich}

The experimental apparatus used to detect normal mode frequencies was a DRS Modulus I Resonant Ultrasound Spectrometer, with lead zirconate titanate transducers. This system has previously been used to successfully determine the elastic constants of Sr$_2$RuO$_4$.\cite{Paglione_RUS}

Once compiled, the list of normal mode frequencies was fitted, and appropriate elastic constants obtained. The measured dimensions of the sample were used as initial dimensions for the program input, but as the measuring errors in these values were the largest sources of error in this study, the dimensions were left as free parameters in the fit. Thus, 9 free parameters -- 6 elastic constants and 3 dimensions -- were fitted to the measured frequency data, while quantities such as the mass and density of the sample were held constant.

Two other samples were cut from the same ingot as the RUS sample and polished to facilitate time-of-flight measurements in the [100] and [110] directions. Sample thicknesses were measured to be 0.319(2) cm  and 0.227(2) cm for the samples oriented along the [100] and [110] directions, respectively. In all cases, transducers were bonded to samples using phenyl salicylate and silicone for room and low temperature measurements respectively. Time-of-flight measurements were made at room temperature (300~K) and low temperature (2~K) using 20 and 30 MHz LiNbO$_3$ transducers and a home-built spectrometer,\cite{Lupien_PhD} which utilizes the phase comparison technique at frequencies between 20 and 500 Mhz. This apparatus has previously been used in a successful study of ultrasonic attenuation in Sr$_2$RuO$_4$.\cite{Lupien_PE}

\begin{center}
\begin{figure*}
 \centering
 \includegraphics[width=6.5in]{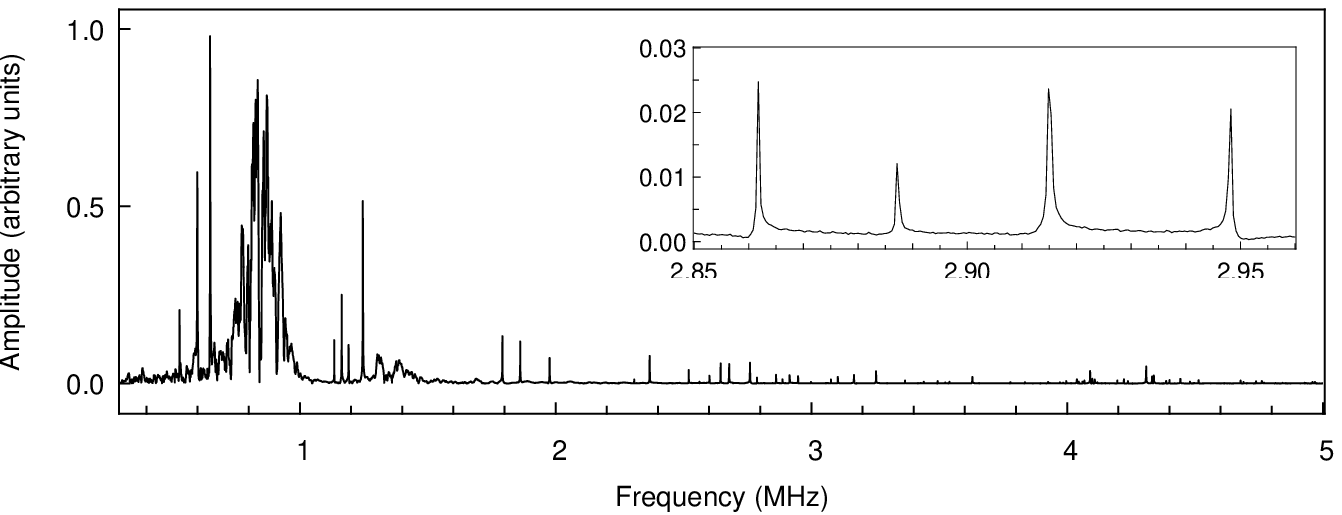}
 \caption{\label{fig:rus_spc} The frequency spectrum measured for YNi$_2$B$_2$C. Inset: a representative portion of this spectrum is expanded, covering a region from 2.85 MHz to 2.96 MHz in detail.}
\end{figure*}
\end{center}

Using the RUS technique, a total of 68 normal modes were measured between 0.5 MHz and 4.5 MHz. The frequencies of these measured modes, along with the frequencies of the calculated modes from the best fitting run and the percent difference between them, are shown in Table~\ref{tab:rus_modes}. The corresponding frequency spectrum is shown in Fig.~\ref{fig:rus_spc}, with a representative portion of the spectrum inset in more detail. In the full-range view, a large amount of noise was measured between approximately 0.7 MHz and 1.1 MHz, and is attributed to frequency characteristics of the measuring apparatus itself. A prototype RUS apparatus, with PVDF (polyvinylidene fluoride) film transducers,\cite{Paglione_MSc} was used to verify that no normal mode frequencies of the sample lay in this range. In addition, the fits indicate no missing modes in this region.

\begin{table*}
\caption{\label{tab:rus_modes}Comparison between experimentally measured normal mode frequencies and normal mode frequencies generated by fit.}
\begin{ruledtabular}
\begin{tabular}{cccc|cccc}
\emph{n} & \emph{f$_{meas}$} (MHz) & \emph{f$_{calc}$} (MHz) & $\vert$ error \% $\vert$ & \emph{n} & \emph{f$_{meas}$} (MHz) & \emph{f$_{calc}$} (MHz) & $\vert$ error \% $\vert$\\
\hline
1 & 0.529514 & 0.528532 & 0.19 & 36 & 3.493864 & 3.508879 & 0.43 \\
2\footnotemark[1] & 0.597221 & 0.591763 & 0.91 & 37 & 3.523911 & 3.513904 & 0.28 \\
3 & 0.647704 & 0.646337 & 0.21 & 38 & 3.540339 & 3.557265 & 0.48 \\
4 & 1.135447 & 1.136584 & 0.10 & 39 & 3.677818 & 3.647198 & 0.83 \\
5 & 1.163147 & 1.162634 & 0.04 & 40 & 3.728029 & 3.719723 & 0.22 \\
6 & 1.190192 & 1.184750 & 0.46 & 41 & 3.778609 & 3.790058 & 0.30 \\
7 & 1.245350 & 1.244220 & 0.09 & 42\footnotemark[2] & - & 3.792207 & - \\
8 & 1.774185 & 1.779893 & 0.32 & 43 & 3.836754 & 3.858453 & 0.57 \\
9 & 1.791589 & 1.784886 & 0.37 & 44 & 3.864763 & 3.861419 & 0.09 \\
10 & 1.860806 & 1.859335 & 0.08 & 45 & 3.902221 & 3.899468 & 0.07 \\
11 & 1.976029 & 1.972153 & 0.20 & 46 & 3.926129 & 3.926370 & 0.01 \\
12 & 2.246914 & 2.253101 & 0.28 & 47 & 3.935864 & 3.935392 & 0.01 \\
13 & 2.307146 & 2.307743 & 0.03 & 48 & 3.971114 & 3.966542 & 0.12 \\
14 & 2.368161 & 2.364670 & 0.15 & 49 & 3.977251 & 3.978405 & 0.03 \\
15 & 2.520553 & 2.520366 & 0.01 & 50 & 3.995687 & 3.988349 & 0.18 \\
16 & 2.561937 & 2.563005 & 0.04 & 51\footnotemark[2] & - & 4.023993 & - \\
17 & 2.601103 & 2.601906 & 0.03 & 52 & 4.039216 & 4.039890 & 0.02 \\
18 & 2.645586 & 2.649300 & 0.14 & 53 & 4.046316 & 4.042717 & 0.09 \\
19 & 2.678345 & 2.663728 & 0.55 & 54 & 4.060185 & 4.053374 & 0.17 \\
20 & 2.688584 & 2.676061 & 0.47 & 55 & 4.067538 & 4.057332 & 0.25 \\
21 & 2.760655 & 2.737530 & 0.84 & 56 & 4.090402 & 4.081257 & 0.22 \\
22 & 2.787477 & 2.789393 & 0.07 & 57 & 4.107759 & 4.082228 & 0.62 \\
23 & 2.848279 & 2.849765 & 0.05 & 58 & 4.119079 & 4.139375 & 0.49 \\
24 & 2.861686 & 2.855943 & 0.20 & 59 & 4.148179 & 4.147882 & 0.01 \\
25 & 2.886700 & 2.889564 & 0.10 & 60 & 4.195629 & 4.176098 & 0.47 \\
26 & 2.915996 & 2.901747 & 0.49 & 61 & 4.221953 & 4.211651 & 0.24 \\
27 & 2.947486 & 2.950772 & 0.11 & 62 & 4.238927 & 4.244522 & 0.13 \\
28 & 2.997868 & 3.004079 & 0.21 & 63 & 4.308970 & 4.332035 & 0.54 \\
29 & 3.076659 & 3.078714 & 0.07 & 64 & 4.332351 & 4.343469 & 0.26 \\
30 & 3.103831 & 3.108858 & 0.16 & 65 & 4.338505 & 4.366086 & 0.64 \\
31 & 3.166186 & 3.178981 & 0.40 & 66 & 4.386658 & 4.378300 & 0.19 \\
32 & 3.251992 & 3.260277 & 0.25 & 67 & 4.399958 & 4.389225 & 0.24 \\
33 & 3.365420 & 3.376704 & 0.34 & 68 & 4.442720 & 4.456998 & 0.32 \\
34 & 3.426382 & 3.440963 & 0.43 & 69 & 4.480628 & 4.504514 & 0.53 \\
35 & 3.439616 & 3.444319 & 0.14 & 70 & 4.494579 & 4.514557 & 0.44 \\
\multicolumn{4}{c|}{\small\sl continued\ldots} & 71 & 4.514518 & 4.520470 & 0.13 \\
\end{tabular}
\end{ruledtabular}
\footnotetext[1]{\mbox{not weighted in the fit, as discussed in the text.}}
\footnotetext[2]{\mbox{missing modes, not detected by the apparatus as explained in the text.}}
\end{table*}

The best fit to the data had a root-mean-square error between measured and calculated frequencies of 0.38\%, indicative of a good fit. The fitted dimensions were 0.0636(6) cm $\times$ 0.0878(8) cm $\times$ 0.236(2) cm, which agree with measurements, also providing confirmation of a good fit. Note that only two modes were not detected, but were determined to be existent through the fitting procedure. Also, the second mode was not weighted in the fit, a common practice among the first 1-2 modes in RUS measurements.\cite{Migliori}

The elastic constants determined via the best fit are listed in Table~\ref{tab:constants}. Error values are obtained from the largest of either an estimation of the quality of the fit, as generated by the fitting program, or from the variation in \emph{C}$_{ij}$ values over the range of densities allowed by the error in the measured dimensions. Specifically, the errors in \emph{C}$_{44}$ and all three fitted dimensions come from the error in density, while the rest come from quality-of-fit estimates.

\begin{figure}
 \includegraphics[width=3.2in]{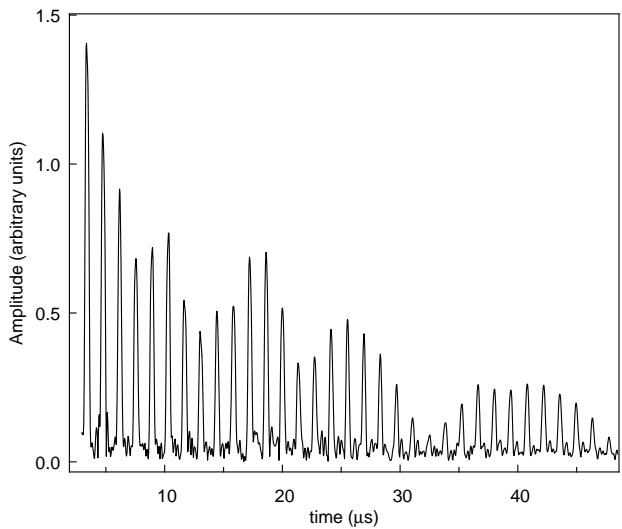}
 \caption{\label{fig:p_echo} Example of raw time-of-flight data, in the case of a low-temperature (2~K) measurement of v$_{[110/T001]}$.}
\end{figure}

Also included in Table~\ref{tab:constants} are the elastic constants determined from three different sources using the time-of-flight technique: room temperature measurements from this study, low-temperature measurements done by our group\cite{Paglione_2K} and low-temperature measurements by Isida \emph{et al.}\cite{Isida} The elastic constants from our room temperature and low-temperature time-of-flight measurements were calculated from velocity data as discussed above.\cite{Truell} An example of the raw time-of-flight data used to obtain velocities is shown in Fig.~\ref{fig:p_echo}. The time delay between each echo is equivalent to the total travel time of an ultrasound pulse going through the sample and back; when taken with the thickness of the sample this yields the velocity of sound for that particular direction of propagation/polarization combination.

The measured and calculated sound velocities with the appropriate polarization and direction are listed in Table~\ref{tab:velocities}. Note that whenever the density of YNi$_2$B$_2$C was required in a calculation, $\rho = 6.3(1)$ g/cm$^3$ was used for measurements performed on our specimen, and $\rho = 6.05$ g/cm$^3$ was used for the measurements by Isida \emph{et al.}

\begin{table*}[!hbt]
\caption{\label{tab:constants}Elastic constants of single crystal YNi$_2$B$_2$C, in units of 10$^{12}$ dynes/cm$^2$.}
\begin{ruledtabular}
\begin{tabular}{cccccccc}
Method &\emph{C}$_{11}$ & \emph{C}$_{33}$ & \emph{C}$_{13}$ & \emph{C}$_{12}$ & \emph{C}$_{44}$ & \emph{C}$_{66}$ & Temperature (K) \\
\hline
RUS & 2.94(6) & 2.61(5) & 1.25(6) & 1.57(7) & 0.644(4) & 1.42(1) & 300 \\
time-of-flight & 2.84(7) & & & 1.45(7) & 0.67(1) & 1.43(3) & 300 \\
time-of-flight & 2.9(2) & & & 1.49(8) & 0.67(4) & 1.3(2) & 2 \\
time-of-flight\footnotemark[1] & 2.2(2) & 2.1(2) & & 1.0(2)\footnotemark[2] & 0.54(5) & 1.3(1) & 14.2 \\
\end{tabular}
\end{ruledtabular}
\footnotetext[1]{\mbox{from Ref.~\onlinecite{Isida}.}}
\footnotetext[2]{\mbox{calculated from the quoted values of \emph{C}$_{11}$ and (\emph{C}$_{11}$ - \emph{C}$_{12}$)/2 in Ref.~\onlinecite{Isida}.}}
\end{table*}

\begin{table*}[!hbt]
\caption{\label{tab:velocities}Sound velocities in YNi$_2$B$_2$C, in units of km/s.}
\begin{ruledtabular}
\begin{tabular}{ccccccccc}
Method & v$_{[100/L]}$ & v$_{[001/L]}$ & v$_{[110/T001]}$, v$_{[100/T001]}$ & v$_{[100/T010]}$ & v$_{[110/T1\bar{1}0]}$ & v$_{[110/L]}$ & Temperature (K) \\
\hline
RUS & 6.8(2) & 6.4(2) & 3.20(5) & 4.74(8) & 3.3(3) & 7.6(2) & 300 \\
time-of-flight & 6.72(6) & & 3.26(2) & 4.76(4) & 3.36(2) & 7.66(4) & 300 \\
time-of-flight & 6.8(3) & & 3.3(2) & 4.5(6) & 3.3(2) & 7.4(4) & 2 \\
time-of-flight\footnotemark[1] & 6.0(3) & 5.9(3) & 3.0(1) & 4.7(2) & 3.2(2) & 6.9(7) & 14.2 \\
\end{tabular}
\end{ruledtabular}
\footnotetext[1]{\mbox{from Ref.~\onlinecite{Isida}.}}
\end{table*}

Upon inspection of the data shown in Tables~\ref{tab:constants} and ~\ref{tab:velocities}, it is immediately apparent that while the values obtained via both the RUS and time-of-flight methods of our group agree within error, the values published by Isida \emph{et al.}\cite{Isida}, with the exception of \emph{C}$_{66}$, are between 16\% and 36\% lower than the RUS results and both sets of time-of-flight results from our group. Reasons for this disagreement are not clear, but agreement of RUS and time-of-flight measurements on samples cut from the same specimen may suggest the significance of differences in sample composition between the two groups.

The RUS values of the elastic constants give a bulk modulus for YNi$_2$B$_2$C of 1.8(1)$\times$10$^{12}$ dynes/cm$^2$. This is in the range of the isothermal bulk modulus values of 1.2(1)$\times$10$^{12}$ dynes/cm$^2$ and 2.00$\times$10$^{12}$ dynes/cm$^2$ previously measured for this material by x-ray diffraction,\cite{Bud'ko,Meenakshi} which gives us added confidence in our results. Furthermore, the Debye temperature was calculated via Eq.~\ref{1:2}, yielding a value of 525(10)~K, which falls within the wide range of published values of the Debye temperature for YNi$_2$B$_2$C: 310(20)~K\cite{Andreone} from a resistivity measurement, and 415~K,\cite{Godart} 489(5)~K\cite{Movshovich} and 537~K\cite{Hong} from specific heat measurements.

In conclusion, all six independent elastic constants of YNi$_2$B$_2$C have been determined using the resonant ultrasound spectroscopy method. Good agreement was found between these results and the elastic constants measured using the time-of-flight technique on a sample from the same ingot. A calculation of the bulk modulus and an estimation of the Debye temperature of YNi$_2$B$_2$C were performed with the elastic constants obtained in this study, and the results were found to be consistent with the range of values found in the literature for these quantities.

\begin{acknowledgments}

This work was supported by the Canadian Institute for Advanced Research and funded by NSERC.  ~J.P. also acknowledges the support of the Walter C. Sumner Foundation.

\end{acknowledgments}

\end{document}